\documentclass[graybox]{svmult}
\usepackage{mathptmx} 
\usepackage{graphicx} 
\begin{document}

\title{Modelling Starburst in H{\sc ii} galaxies: From chemical to spectro-photometric evolutionary self-consistent models.}
 \titlerunning{Modelling Starburst in H{\sc ii} galaxies.} 
\author{Mart\' {i}n-Manj\'{o}n, M.L., Moll\' {a}, M., D\' {i}az, A.I. \& Terlevich, R.}
\institute{Mart\' {i}n-Manj\'{o}n, M.L \at Universidad Autónoma de Madrid, Madrid (Spain), \email{mariluz.martin@uam.es}
\and Moll\' {a}, M. \at CIEMAT, Madrid (Spain), 
\and M., D\' {i}az, A.I \at Universidad Autónoma de Madrid, Madrid (Spain), 
\and Terlevich, R. \at INAOE, Puebla (Mexico) 
}
%
%
\maketitle

\abstract*{We have computed a series of realistic and self-consistent models that reproduce the properties of H{\sc ii} galaxies. 
 The emitted spectrum of H{\sc ii} galaxies is reproduced by means of the photoionization code CLOUDY, using as ionizing spectrum the spectral energy distribution of the modelled H{\sc ii} galaxy, calculated using new and updated stellar population synthesis model (PopStar) 
 This, in turn, is calculated according to a star formation history and a metallicity evolution given by a chemical evolution code. Our technique reproduces observed abundances, diagnostic diagrams, colours and equivalent width-­colour relations for local H{\sc ii} galaxies.}

\abstract{We have computed a series of realistic and self-consistent models that reproduce the properties of H{\sc ii} galaxies. 
 The emitted spectrum of H{\sc ii} galaxies is reproduced by means of the photoionization code CLOUDY, using as ionizing spectrum the spectral energy distribution of the modelled H{\sc ii} galaxy, calculated using new and updated stellar population synthesis model (PopStar) 
This, in turn, is calculated according to a star formation history and a metallicity evolution given by a chemical evolution code. Our technique reproduces observed abundances, diagnostic diagrams, colours and equivalent width-­colour relations for local H{\sc ii} galaxies.}

%

\section{The Model.}
\label{sec:1}
H{\sc ii} galaxies are characterized by strong and narrow emission lines and by a low metal 
content, but this does not necessary mean that these galaxies be young systems. The 
current burst of star formation (SF) dominates the spectral energy distribution (SED) even if 
previous stellar populations are present, making difficult to know the star formation history 
(SFH) of the galaxy. We have made a new grid of star-bursting models, based on \cite{mmm08}, using simultaneously the whole information available for the galaxy 
sample: the ionized gas, which defines the present time state of the galaxy, and 
spectrophotometric parameters, related to its SFH. These models imply the combination of 
three different tools: a chemical evolution code, an evolutionary population synthesis 
code and a photo­ionization code, all computed in a self-­consistent way, that is, using the 
same assumptions regarding stellar evolution, model stellar atmospheres and 
nucleosynthesis, and a realistic age-­metallicity relation.


The model consist in a set of successive burst of SF in a region with a total mass of gas of 100 10$^{6}$ M$_{\odot}$. Each model is characterized by the \textbf{initial efficiency}, $\varepsilon$, the percentage of available gas consumed to form stars, and the \textbf{attenuation}: the initial efficiency is attenuated by a constant factor, \textit{k}, in the successive burst following the expression $\psi_{n}= \psi_{0}(k)^{(n-1)}$, where \textit{n} is the number of the current burst.

We present two models with \textbf{k=0.2} and two different initial efficiencies: \textbf{Model 1}, with $\varepsilon$=50$\%$ and \textbf{Model 2} with $\varepsilon$=12$\%$.

With the chemical evolution code, based on \cite{fer94}, we obtain for every 0.7 Myr step the abundances of 15 elements: H, He, C, O, N, Ne, Na, Al, Mg, Si, S, Ca, Ar, Ni, Fe, the star formation rate (SFR) and the corresponding age-metallicity relation, Z(t).


A SED from the Single Stellar Populations (SSPs) library PopStar \cite{mol08}, has been assigned to each time step taking into account the SFH, $\varepsilon$, $\Psi$(t) and Z(t) obtained from the chemical evolution model. When more than one burst take place, the SED is the sum of the SEDs of every stellar generation convolved with the SFH. The final result is the total luminosity at each wavelength for each time step, L$_{\lambda}$ (t): the SED corresponding to the whole stellar population, including the ionizing continuum of the last formed stellar generation.


The gas is ionized by the massive stars belonging to the current burst of SF. We use CLOUDY \cite{fer98} in order to obtain emission lines. Each burst photoionization model is characterized by a radius R, calculated according to the mechanical energy output of the massive star winds and SNeI explosions, a constant gas density n$_{H}$=100 cm$^{-3}$, an ionization parameter U = Q(H)/4$\pi$cn$_{H}$R$^{2}$, where \textit{Q(H)} is the number of ionizing photons, and the chemical abundances obtained from the chemical evolution code.

\section{Results.}

\begin{figure}[!ht]
\sidecaption[t]
\includegraphics[scale=.25,clip]{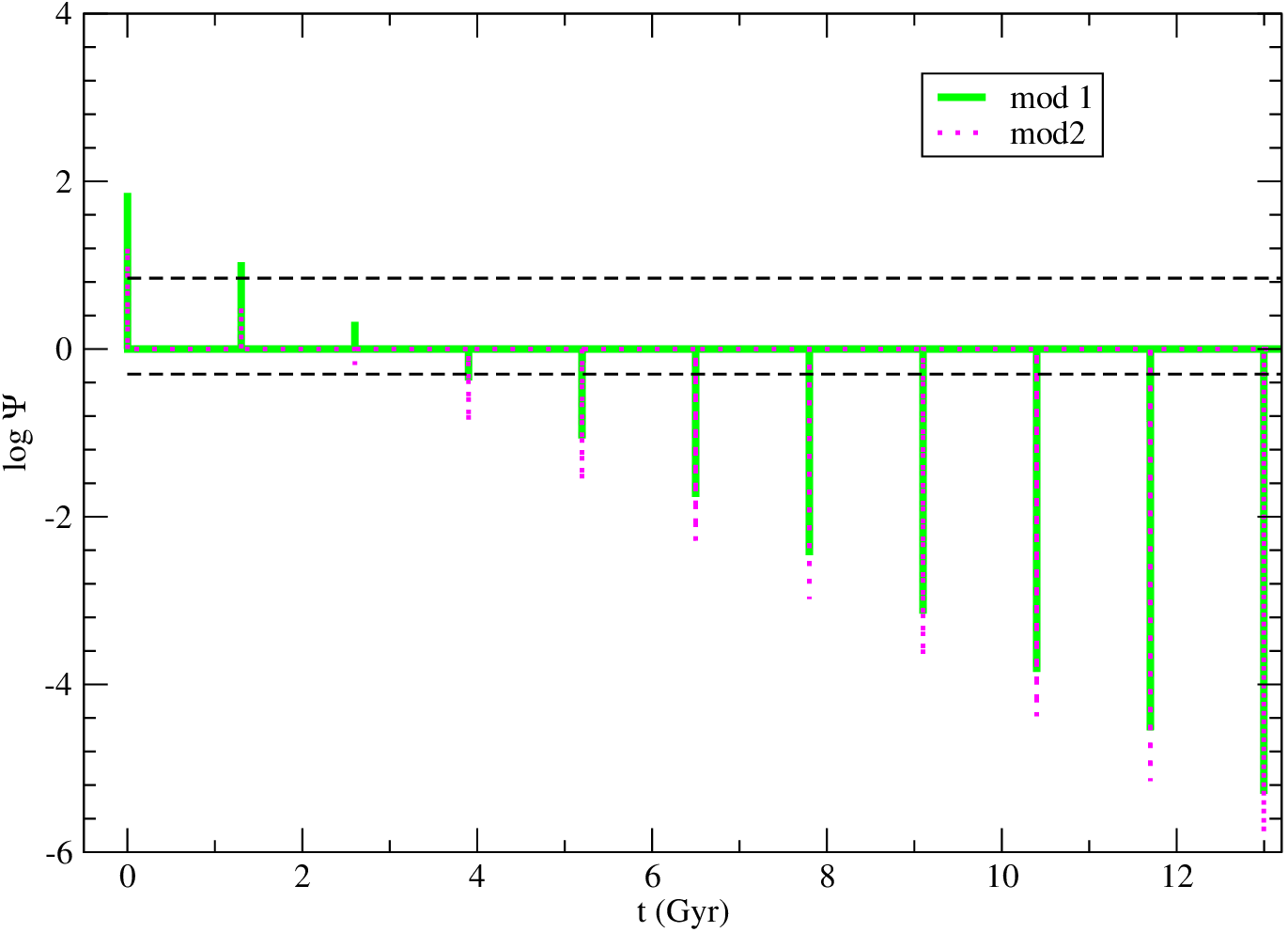}
\includegraphics[scale=.25,clip]{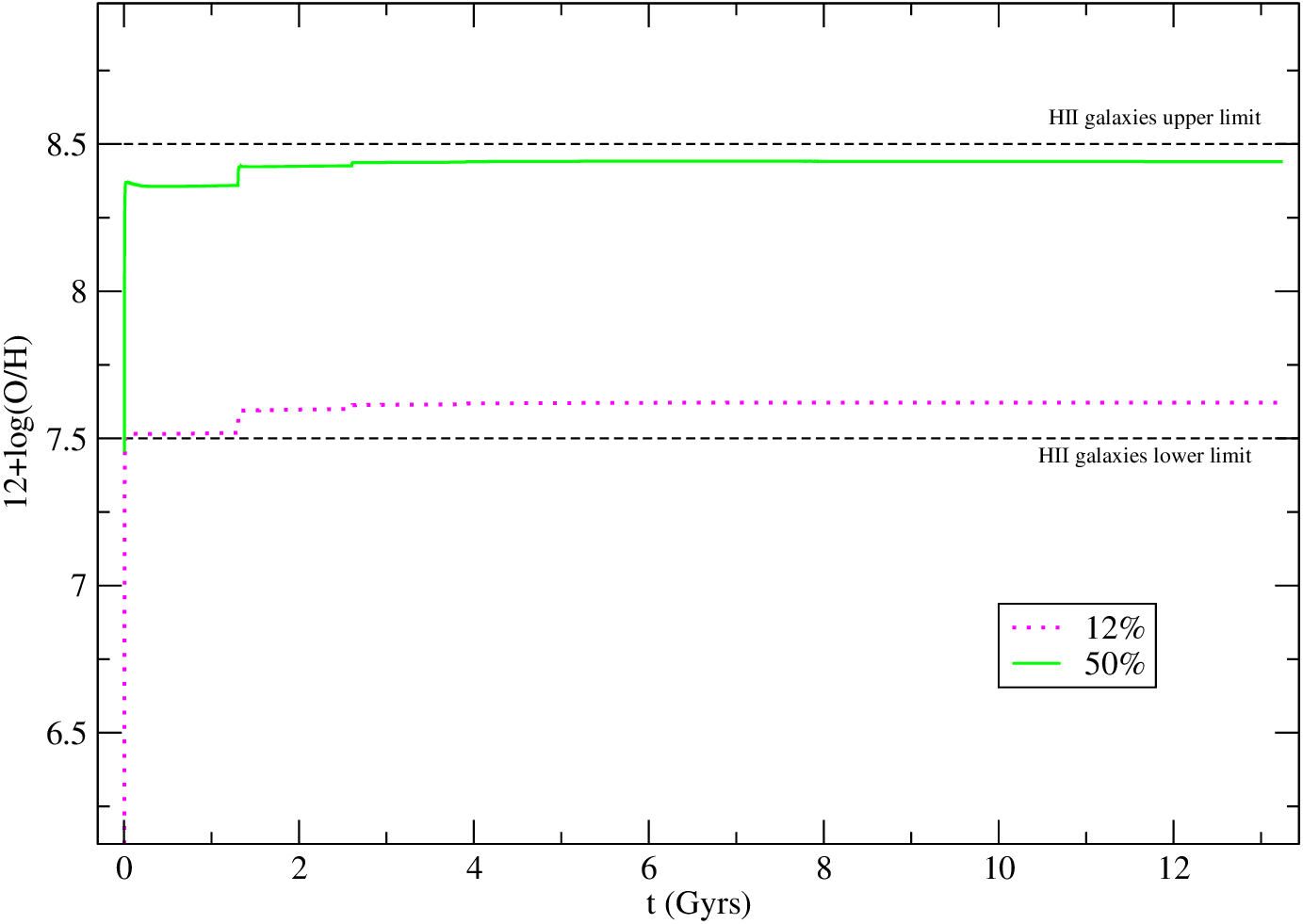}
%
%
\caption{SFR along 13.2 Gyr of Model 1 and Model 2 (left). Evolution of the oxygen abundance for both models (right). Black dashed lines corresponds to the observed limits for H{\sc ii} galaxies \cite{hoy06}.}
\label{fig:1} 
\end{figure}

In fig.\ref{fig:1} (left) we represent the SFR of Model 1 and Model 2, and the evolution of the oxygen abundance along 13.2 Gyr (right). The first burst is strong, while subsequent are less intense due to the strong attenuation and the decrease of the available gas to form stars. We see in fig. \ref{fig:1} (right) that the efficiencies of the Model 1 and Model 2 are the upper and lower limits, respectively, for the oxygen abundance in H{\sc ii} galaxies. 

\begin{figure}[!ht]
\sidecaption[t]
\includegraphics[scale=.25,clip]{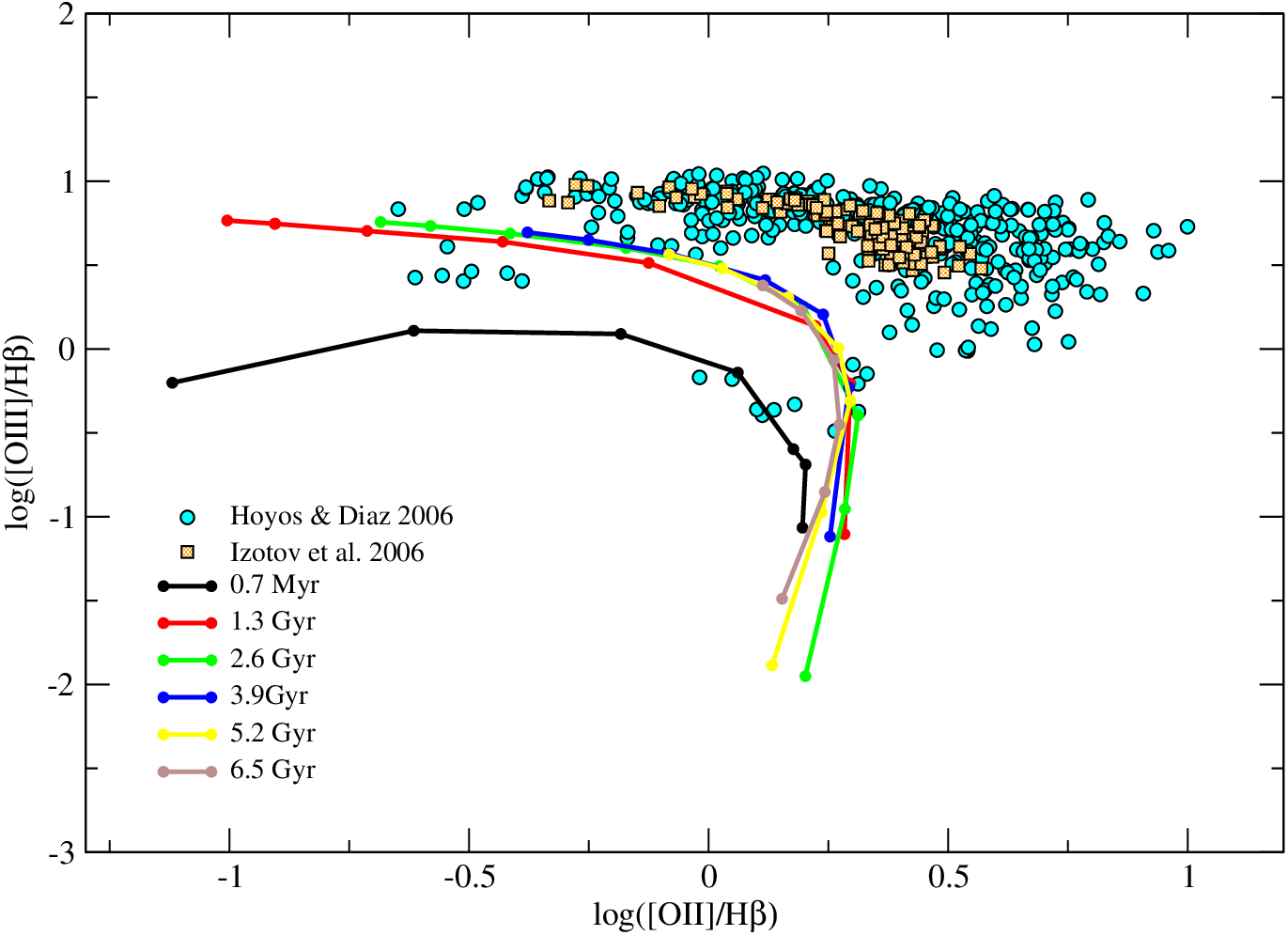}
\includegraphics[scale=.25,clip]{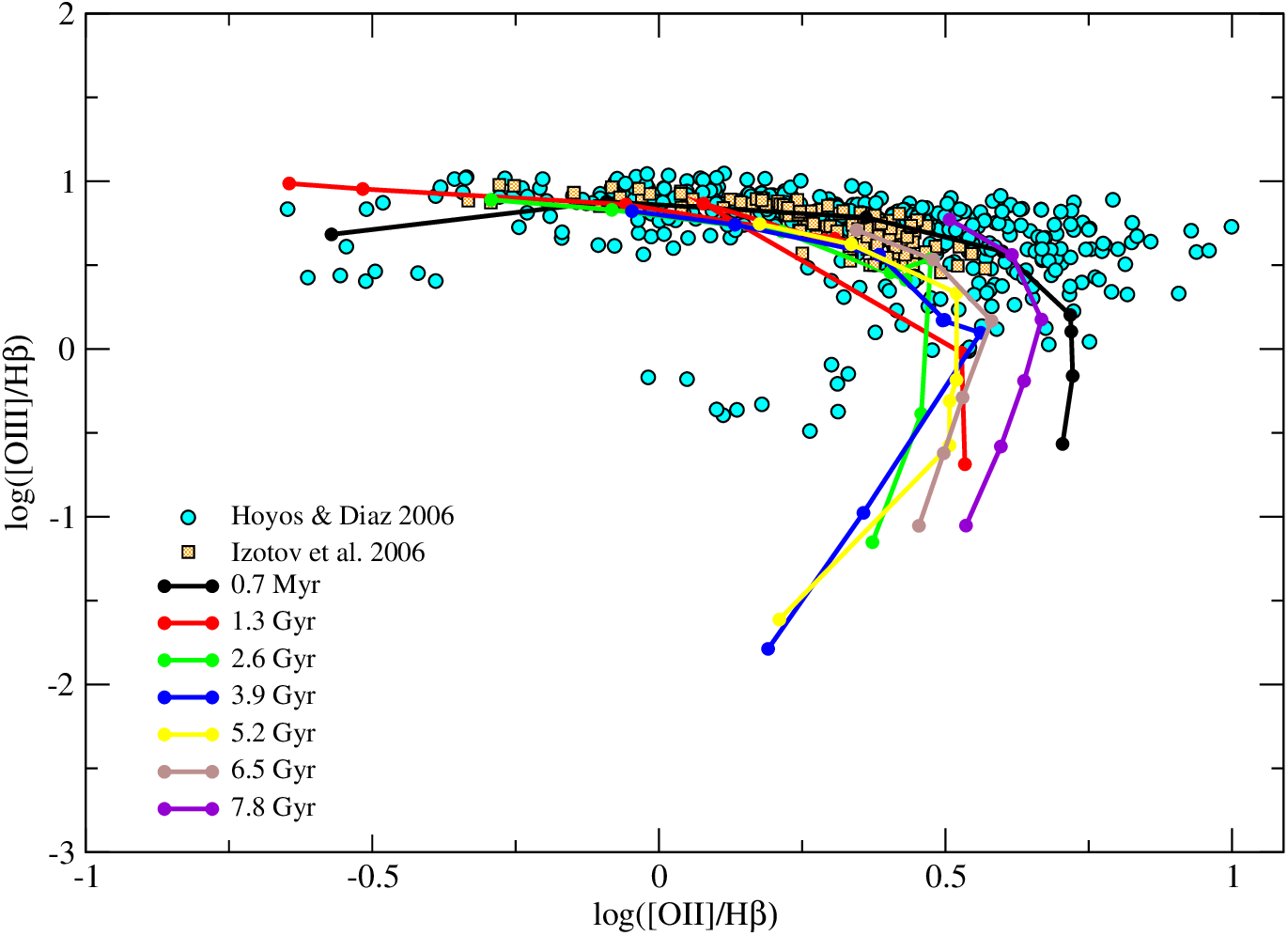}
%
%
\caption{Diagnostic diagrams for low efficiency model (left) and high efficiency model (right)}
\label{fig:2} 
\end{figure}

The properties of the ionized gas, showed as excitation diagnostic diagrams, can be seen in fig.\ref{fig:2}. Model 1 reproduces the more excited and more metallic galaxies, with high [OIII]/H$_{\beta}$ due to its high efficiency of SF, while Model 2 reproduces less metallic galaxies, with high [OIII]/H$_{\beta}$ and low [OII]/H$_{\beta}$ ratios.

\begin{figure}[!ht]
\sidecaption[t]
\includegraphics[scale=.25,clip]{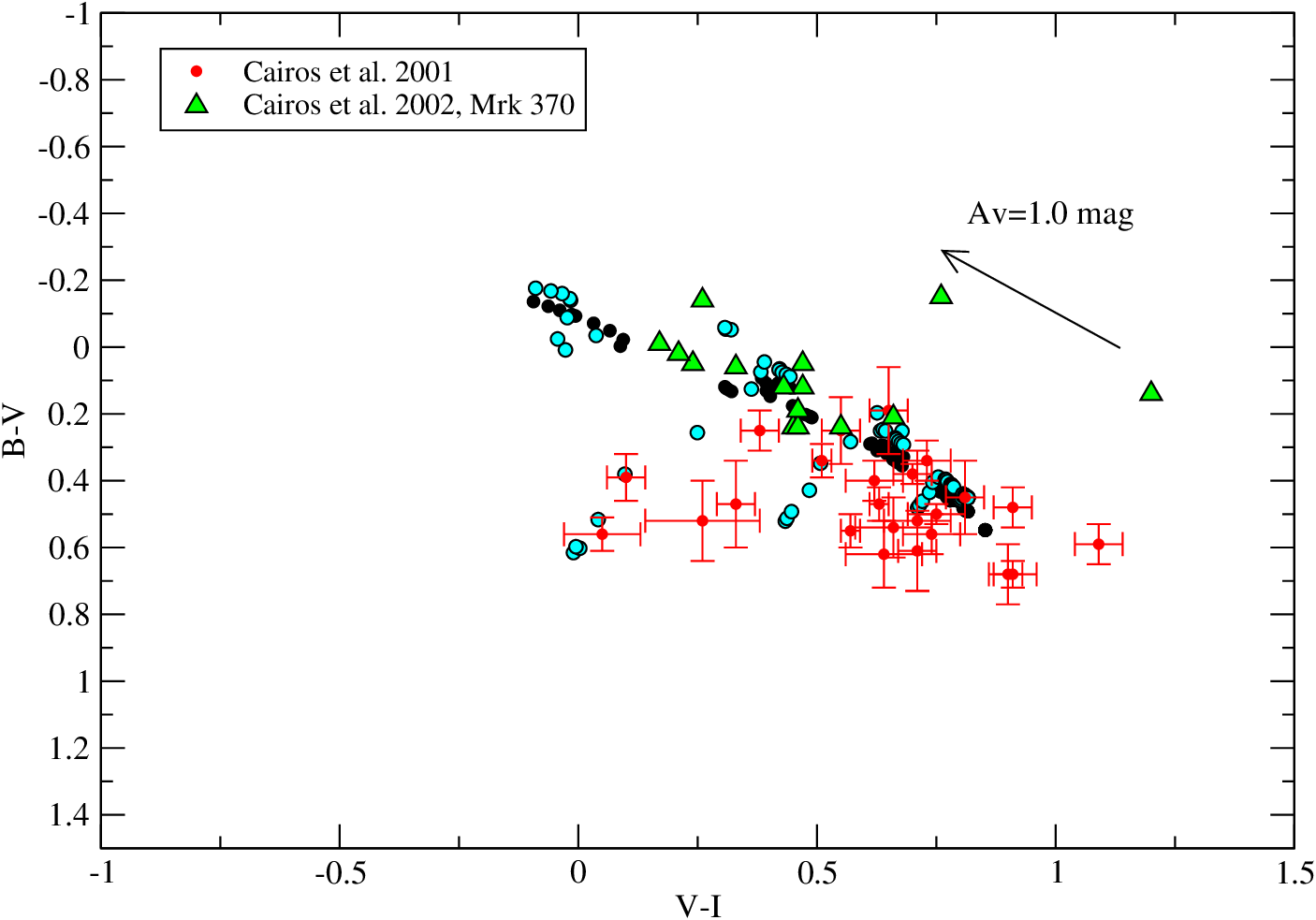}
\includegraphics[scale=.25,clip]{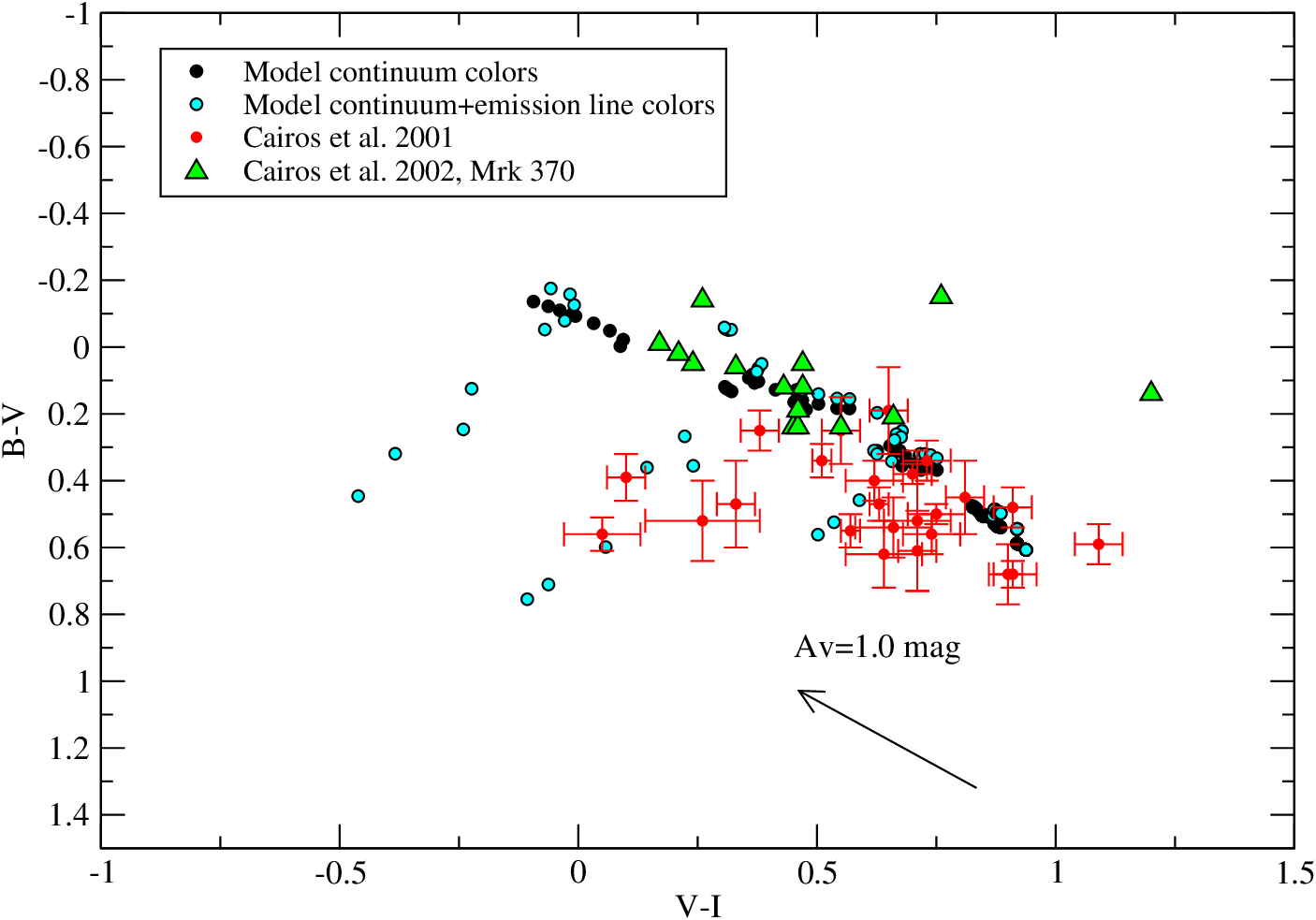}
%
%
\caption{Colour-colour diagram for Model 1 (right) and Model 2 (left).}
\label{fig:3} 
\end{figure}

The broad band colours have been computed including the contribution by the stronger emission lines. These are the colours which are readily observable though integrated photometry (fig.\ref{fig:3}).

\begin{figure}[!ht]
\begin{center}
\includegraphics[scale=.25,clip]{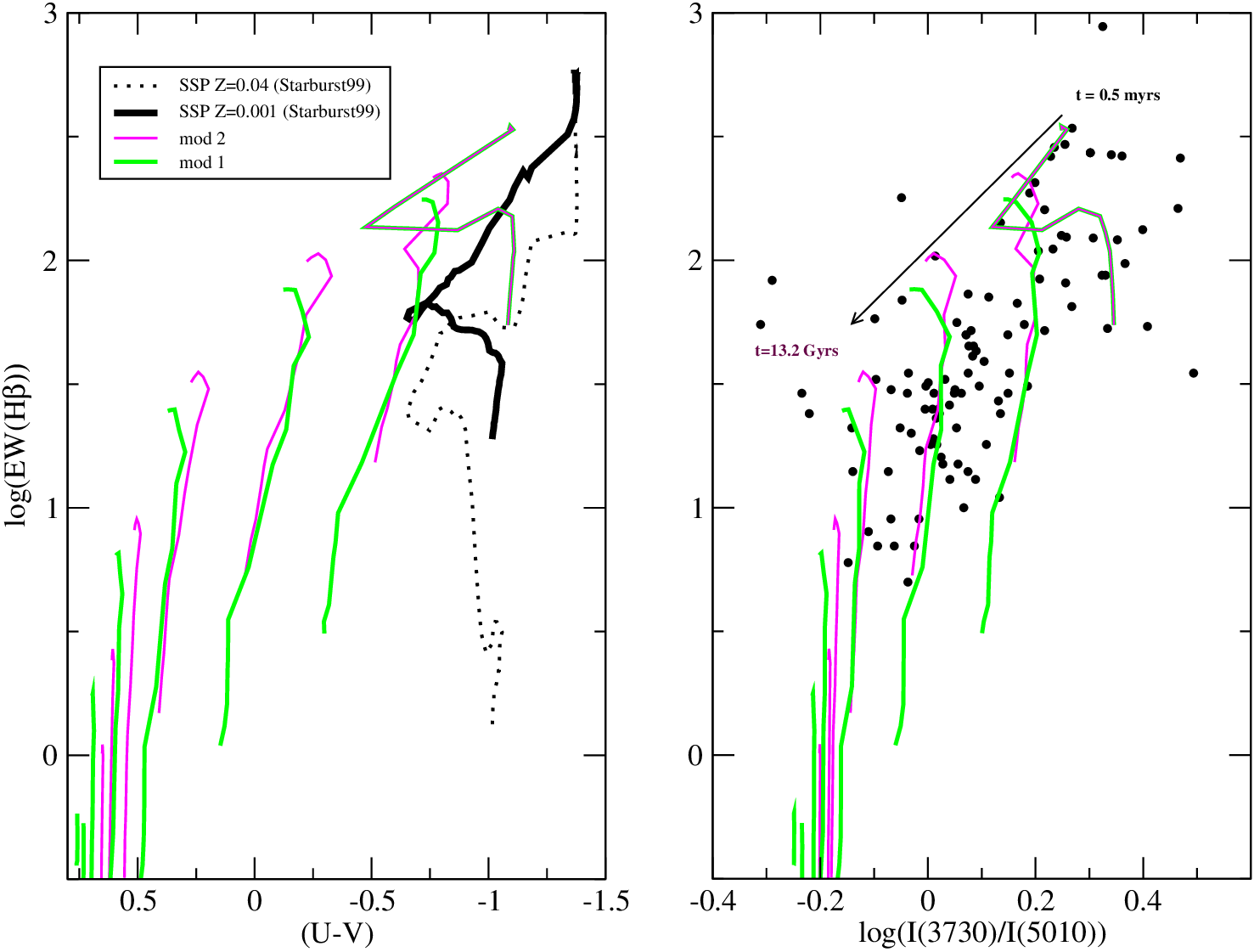}
%
%
\caption{(left) EW(H$_{\beta}$) vs (U-V), compared with two SSPs from Starburst99 \cite{stb99}; a very low metallicity SSP (black solid line) and a high metallicity SSP (dotted line). The contribution of the underlying continuum make the colour to be redder and reproduce observational trend of H{\sc ii} galaxies, as we see on the right: EW(H$_{\beta}$) vs a pseudo-colour equivalent to U-V.}
\label{fig:4} 
\end{center}
\end{figure}

Most H{\sc ii} galaxies have EW(H$_{\beta}$) values lower than 150 $\AA$, which indicates the presence of an old non-ionizing population. According to the models, these values can not be produced by just one generation of stars (a SSP), and it is necessary to invoke an underlying population, of at least 1.3 Gyr, to contribute to the continuum. EW(H$_{\beta}$) vs. (U-V) along the evolution of the galaxy for the stellar bursts is shown in fig.\ref{fig:4}. The trend of the data is reproduced by our models and not by SSP's since in that case the colours are bluer than observed. In order to decrease EW(H$_{\beta}$) and to obtain redder colours, a more metal-rich SSP might be selected (dotted line at the left panel), but such high abundance does not reproduce the observations (fig. \ref{fig:1}).

In summary, our models are able to produce redder colours with the contribution of previous stellar generations and reproduce better the trend of H{\sc ii} galaxies observations, as well as abundances, colours and emission lines.


%



\begin{thebibliography}{}

\bibitem {cai012}
{Cair{\'o}s} L.~M., {Caon} N., {V{\'{\i}}lchez} J.~M.,
et~al. 2001, ApJS, 136, 393

\bibitem{fer98}
{Ferland} G.~J., {Korista} K.~T., {Verner} D.~A., et.~al. 1998, PASP, 110, 761




\bibitem {fer94} {Ferrini} F.,
{Moll\'{a}} M., {Pardi} M.~C., {D{\'{\i}}az} A.~I., 1994, ApJ, 427,
745

\bibitem {hoy06} 
{Hoyos} C., {D{\'{\i}}az} A.~I., 2006,
MNRAS, 365, 454


\bibitem{stb99}
{Leitherer} C., {Schaerer} D., {Goldader} J. D., {Delgado} R. ~M. ~G., 
{Robert} C., {Kune} D. ~F., {de Mello} D. ~F., {Devost} D., {Heckman} 
T. ~M., 1999, ApJS, 123, 3

\bibitem {mmm08}
Mart\'{\i}n-Manj\'{o}n, M.L.,Moll\'{a}, M., D\'{\i}az, A.I. 
\& Terlevich, R. 2008, MNRAS, 

\bibitem {mol08} 
Moll{\'a} M., Garc{\'{\i}}a-Vargas M.~L. \& Bressan, A. 2008, in preparation.

\end{thebibliography}
\end{document}